\documentclass[aps,a4paper,twocolumn,showpacs]{revtex4}

\usepackage{graphicx}

\begin{document}

\newcommand{\be}   {\begin{equation}}
\newcommand{\ee}   {\end{equation}}
\newcommand{\ba}   {\begin{eqnarray}}
\newcommand{\ea}   {\end{eqnarray}}
\newcommand{\tr}   {{\rm tr}}

\title{Transfer Matrices and Circuit Representation 
       for the \\ Semiclassical Traces of the Baker Map} 

\author{Gabriel G. Carlo}
\email{carlo@tandar.cnea.gov.ar}
\affiliation{ Departamento de F\'{\i}sica, 
              Comisi\'on Nacional de Energ\'{\i}a 
              At\'omica, 
              Avenida del Libertador 8250, 
             (1429) Buenos Aires, Argentina}

\author{Ra\'ul O. Vallejos}
\email{vallejos@cbpf.br}

\author{R\^omulo F. Abreu}
\email{romulo@cbpf.br}
\affiliation{ Centro Brasileiro de Pesquisas F\'{\i}sicas (CBPF), 
              Rua Dr.~Xavier Sigaud 150, 
              22290-180 Rio de Janeiro, Brazil}

\date{\today}

\begin{abstract}
Because of a formal equivalence with the partition function 
of an Ising chain, the semiclassical traces of the quantum 
baker map can be calculated using the transfer-matrix
method.
We analyze the transfer matrices associated with the baker
map and the symmetry-reflected baker map
(the latter happens to be unitary but the former is not).
In both cases simple quantum-circuit representations are
obtained, which exhibit the typical structure of qubit 
quantum bakers.
In the case of the baker map it is shown that nonunitarity
is restricted to a one-qubit operator (close to a Hadamard 
gate for some parameter values).  
In a suitable continuum limit we recover the already known 
infinite-dimensional transfer-operator.
%
%
We devise truncation schemes allowing the calculation of 
long-time traces in regimes where the direct 
summation of Gutzwiller's formula is impossible.
Some aspects of the long-time divergence
of the semiclassical traces are also discussed.
\end{abstract}




\maketitle

\section{Introduction}
\label{sec1}

The quantum baker's map \cite{balazs87} is a very useful 
test-bench for investigating quantum-classical correspondence 
issues in a variety of settings.
Conceived as a model for studying the semiclassical
limit of closed chaotic systems, it was first applied to
analyze the random-matrix conjecture, the scarring phenomenon,
Gutwiller's trace formula and the long-time validity of 
semiclassical approximations 
(see, e.g., \cite{balazs89, saraceno90, ozorio91, saraceno94, 
dittes94, heller96}).
In the last years ``open" quantum bakers were constructed
and employed for studying semiclassical aspects of the 
scattering problem \cite{openbaker}, e.g., the fractal Weyl 
law for the distribution of resonances \cite{schomerus04}.

The quantum baker also appeared in a variety of problems 
of Quantum Information, Quantum Computation and Quantum Open 
Systems.
It was noted that the quantum baker could be efficiently 
realized in terms of quantum gates \cite{schack98}.
A three qubit Nuclear Magnetic Resonance experiment was 
proposed \cite{brun99} and then implemented (with some 
simplifications) \cite{weinstein02}.

On the theoretical side, Schack and Caves \cite{schack00} 
showed that the Balazs-Voros-Saraceno quantum baker 
\cite{balazs89, saraceno90} can be seen as a shift 
on a string of quantum bits (in full analogy with the 
classical case), thus taking an important step towards 
generalizing the method of symbolic dynamics to the quantum 
case. 
At the same time, their circuit representation led naturally
to a family of alternative quantizations.
This family of bakers was the subject of several studies
\cite{soklakov00,tracy02,scott03}. 
The ability of the baker family to generate entanglement
was first studied by Scott and Caves \cite{scott03}
(see also \cite{abreu06}). 
Decoherent variants of the baker map were constructed 
by including mechanisms of dissipation and/or diffusion 
\cite{lozinski02, soklakov02, bianucci02}.

From the point of view of the structure of the quantum 
baker map, important results were recently obtained by 
Ermann and Saraceno \cite{ermann06}, who, building upon 
previous work by Lakshminarayan and Meenakshisundaram 
\cite{laksh05}, generalized the Schack-Caves family, 
and showed that all quantum bakers are perturbations of 
a common kernel (the ``essential" baker).

The present paper focuses on an almost unexplored aspect 
of the semiclassical theory of the quantum baker map. 
Gutzwiller's approximate formula for the traces of the baker 
map is formally equivalent to the partition function of a
finite Ising chain (with exponentially decaying interactions
and imaginary temperature).
Thus, it can be evaluated with the standard transfer-matrix 
method.
We have studied extensively the transfer matrices 
associated with the semiclassical traces of the baker 
map and the symmetry-reflected baker map.
Our most remarkable finding is the existence of a qubit 
structure hidden in the semiclassical traces: 
the transfer matrices admit a quantum-circuit representation
that is very similar to that found by Schack-Caves for
the quantum baker.

We know of two studies of the baker map which applied the
transfer matrix method 
(to the semiclassical long-time propagation of wave-packets 
\cite{heller96}, to the analysis of periodic-orbit action 
correlations \cite{smilansky03}).  
Different in spirit, the present work aims at analyzing 
the transfer-matrix method in itself.  


The paper is organized as follows.
First we discuss the general structure of the transfer 
matrices for the baker map, exhibiting 
their quantum-circuit representation.
Though the transfer matrices are not unitary, nonunitarity 
is restricted to a one-qubit ``gate". 
Except for this fact, the transfer matrices resemble
qubit quantum bakers (Sec.~\ref{sec2}).

Section~\ref{sec3} is devoted to spectral properties.
We show that all eigenvalues lie inside the unit circle 
in the complex plane, or very close to it. 
The number of eigenvalues close to the unit circle
coincides approximately with the quantum dimension 
(for a suitable parameter range). 
This situation is very similar to that found in the
infinite-dimensional transfer-operator method 
\cite{dittes94,tanner99}.
Indeed we show that the transfer operator arises 
as a suitable continuum limit of the transfer matrix.

In Sec.~\ref{sec4} we exhibit truncation schemes which
permit the calculation of long-time traces.

Some aspects of the long-time divergence of the 
semiclassical traces are assessed in Sec.~\ref{sec5}.

The baker map possesses a spatial symmetry.
If one is interested in separating the spectrum/traces
into symmetry classes, then one must 
consider also the symmetry-related transfer matrices.
Though similar to the matrices of Sec.~\ref{sec2} in
many respects, these matrices happen to be exactly 
unitary.
They are studied in Sec.~\ref{sec6}.

Concluding remarks are presented in Sec.~\ref{sec7}.

\section{Baker map: Transfer matrix}
\label{sec2}

In order to make the paper self-contained we start by
summarizing some basic information about the baker map.
 
\begin{figure}[htp]
\hspace{0.0cm}
\includegraphics[angle=0.0, width=6cm]{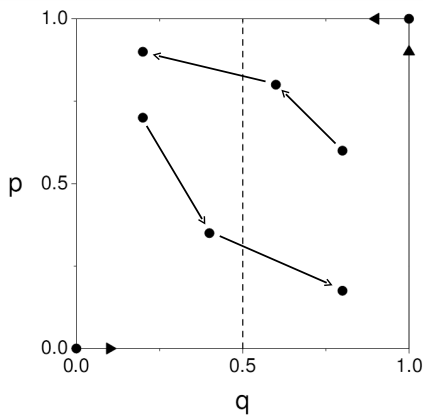}
\caption{The classical baker map. 
The dynamics is controlled by the fixed points
$(0,0)$ (for $q \le 0.5$) and $(1,1)$ ($q>0.5$).
The action of the map is to compress/stretch 
the coordinates $p/q$ relative to the fixed points
by a factor of two. Some examples are depicted.}
\label{fig1}
\end{figure}

The dynamics of the classical baker map is schematically 
depicted in Fig.~\ref{fig1}.
%
%
%
%
It acts on the unit square piecewise-linearly. 
Points with $q < 0.5$ are governed by the hyperbolic point 
at the origin. 
If $q \ge 0.5$ the fixed point at $(1,1)$ rules.
In any case the coordinates $p/q$ relative to the
fixed point are compressed/stretched by a factor of 
two:
\ba
p^\prime & = & \frac{1}{2} \left( p + \epsilon \right) \, , \\
q^\prime & = &               2 \, q - \epsilon          \, ,
\ea
where $\epsilon = [2q]$, the integer part of $2q$.

The quantum analogue of the classical mapping is a unitary 
operator acting on an even-dimensional Hilbert space
\cite{balazs87,balazs89}.
In the $q$-representation its matrix reads \cite{saraceno90}:
\be
B= G_{N}^{-1} \left( \begin{array}{cc}
                           G_{N/2} &      0            \\
                              0    &   G_{N/2}
                          \end{array}
                  \right) \, ,
\label{bakerqq}
\ee
where $G_N$ is the $N$-dimensional antiperiodic Fourier
matrix:
\be
(G_{N})_{mn}=
\frac{1}{\sqrt N} \, e^{-2\pi i (n+1/2)(m+1/2)/N} \, ,
\ee
with $0 \le m,n \le N-1$.
For this abstract system the Planck constant coincides with
the inverse of the dimension, i.e., $\hbar = 1/(2 \pi N)$.

Our main concern are the traces of the quantum baker, 
${\rm tr} \, B^L$, for $L=1,2,\ldots$.
In the semiclassical regime of $N$ large enough (for a given 
$L$) the traces can be approximated by the Gutzwiller formula 
\cite{saraceno94}
\be
{\rm tr} \, B^L \approx
\frac{2^{L/2}}{2^{L}-1} 
\sum_\epsilon  
e^{2\pi i N {\cal S}(\epsilon)} ~,
\label{gutzwiller}
\ee
where the sum runs over all periodic trajectories of length
$L$ of the classical map, indexed by the binary column vectors
$\epsilon=(\epsilon_0,\epsilon_1,\ldots,\epsilon_{L-1})^t$, 
with
$\epsilon_i=0,1$.  
The corresponding actions ${\cal S}(\epsilon)$ are quadratic 
functions of the binary symbols,
\be
{\cal S}(\epsilon)=\epsilon^t A \, \epsilon \; .
\label{action}
\ee
The ``coupling" matrix $A$ is suitably expressed in terms of 
the matrix $S$ of a cyclic shift, 
$S \cdot  (a_0,a_1,\ldots a_{L-1})^t = 
                (a_1,\ldots ,a_{L-1},a_0)^t$ 
\cite{toscano97}:
\be
A = \frac{2^{L-1}}{2^{L}-1}
          \sum_{i=0}^{L-1}
          \frac{S^i}{2^i}  \;.
\label{matrixA}
\ee

The trace formula for the baker map (\ref{gutzwiller}) 
can be derived as follows. 
Write ${\rm tr} \, B^L$ as a sum of products of Fourier 
matrix elements using Eq.~(\ref{bakerqq}).  
Approximate sums by integrals. 
Extend the limits of integration (which in principle 
are finite) to $\pm \infty$. 
The remaining integrals, being Gaussian, can be done 
exactly, giving the Gutzwiller sum \cite{saraceno94}.

Equation~(\ref{gutzwiller}) is a particular case of the
general Gutzwiller trace formula for systems with a chaotic 
classical limit. 
For such systems the trace formula lays a bridge between the 
quantum spectrum and the set of classical periodic orbits.
It constitutes the core of all semiclassical schemes for
relating energy levels to classical information, from the 
pioneer attempts \cite{gutzwiller82, gutzwiller}, to the most 
recent, highly sophisticated developments \cite{haake}.

The baker is very special in that all its periodic orbits,
and properties thereof, are known analytically.
So, in principle, the periodic-orbit sum (\ref{gutzwiller}) 
can be calculated for any $L$. 
However, because of the exponential growth of the number of 
periodic orbits with period, the brute-force summation
is restricted to, say, $L \le 30$.
Remarkably, the method developed by Dittes, Doron and 
Smilansky \cite{dittes94} does not suffer from such a 
limitation.
These authors defined an infinite-dimensional integral
operator,
\be
W(q^\prime,q)=
\frac{1}{\sqrt{2}} 
\left[
\delta\left(q-\frac{q^\prime  }{2} \right)+
\delta\left(q-\frac{q^\prime+1}{2} \right) e^{2\pi i N q}
\right]  \, ,
\label{Wqq}
\ee
$ 0 \le q, q^\prime \le 1$, 
whose traces give exactly the periodic-orbit sum of 
Eq.~(\ref{gutzwiller}).
They showed that ${\rm tr} \, W^L$ can be efficiently calculated 
from the eigenvalues of the matrix of $W$ in the Fourier basis,
after suitable truncation \cite{dittes94,tanner99}.

An alternative to the infinite-dimensional operator method
relies on the formal identification of the periodic-orbit 
sum (\ref{gutzwiller}) with the partition function of an 
Ising chain, for a purely imaginary temperature.
The cyclical nature of the coupling matrix  $A$ says
that this is a circular chain, consisting of $L$ spins.
All spins interact among themselves, but, as interactions
decay exponentially with distance, one may expect some
computational benefit (without much loss of accuracy) 
by truncating the interaction to some given number $r$ of 
closest neighbors, i.e.,
\be
A \approx \frac{2^{L-1}}{2^{L}-1}
                \sum_{i=0}^{r}
                \frac{S^i}{2^i}  \;.
\label{truncation}
\ee
The exact periodic-orbit sum corresponds to setting $r=L-1$
(no truncation at all), but, in principle, one can consider 
any truncation, even to first neighbors ($r=1$).

Once one has established the equivalence between the baker 
periodic-orbit sum and the Ising partition function, the  
transfer-matrix method can be invoked. 
The first-neighbor case is explained in textbooks \cite{reichl}. 
It seems that the many-neighbor case has not been
explicitly worked out in the literature, but some discussions
exist \cite{hints,gutzwiller}.
Anyway, even if not trivial, the generalization 
can be carried out following the spirit of the one-neighbor 
case and, in the case of the baker-Ising, leads to 
\be
\frac{2^{L/2}}{2^{L}-1} 
\sum_{\epsilon}  
e^{2\pi i N {\cal S}(\epsilon)} 
\approx 
\frac{2^L}{2^L-1} 
{\rm tr} \, M^L  \,,
\label{transfer}
\ee
the explicit expression for $M$ being \cite{smilansky03}:
\be
M = \frac{1}{\sqrt{2}}
\left( \begin{array}{cccccccc}
          1 & 1 & 0 & 0 & \cdots & \cdots & 0 & 0  \\
          0 & 0 & 1 & 1 &    0   & \cdots & 0 & 0  \\
   \vdots & \vdots & \vdots & \vdots & 
   \vdots & \ddots & \vdots & \vdots               \\
          0 & 0 & 0 & 0 & \cdots & \cdots & 1 & 1  \\
      a_0 & a_1 & 0 & 0 & \cdots & \cdots & 0 & 0  \\
      0 & 0 & a_2 & a_3 & \cdots & \cdots & 0 & 0  \\
   \vdots & \vdots & 0 & 0 & \cdots & 
   \ddots & \vdots & \vdots  \\
          0 & \cdots & 0 & 0 & \cdots & 
              \cdots & a_{2^r-2} & a_{2^r-1} 
       \end{array} \right) \,.
\label{transfer2}
\ee
The elements $a_k$ are given by
\be
a_k=\exp \left[ i \alpha
\left( 1+ \frac{k}{2^r} \right) \right] \,,
\ee
with $0 \le k \le 2^r-1$, and
\be
\alpha= \frac{\pi N \, 2^L }{2^L-1}  \,.
\ee
Note that $M$ is a complex matrix depending on the three 
parameters $(N,L,r)$. 
Its dimension is $2^r \times 2^r$, with $r \le L-1$. 
When $r=L-1$ one recovers the exact semiclassical traces,
i.e., Eq.~(\ref{transfer}) becomes an equality.

A prefactor $1/\sqrt{2}$ has been absorbed into the 
definition of $M$ because in this way most of the 
spectrum of $M$ lies close to the unit circle or inside 
it (see Sec.~\ref{sec3} below).
So, the transfer matrix  $M$ becomes qualitatively
similar to the semiclassical transfer operator of 
Refs.~\cite{dittes94,tanner99} 
(we come back to this point later).

The transfer matrix approach transforms the calculation of 
the Gutzwiller sum into the problem of obtaining the
trace of the $L$-th power of the finite matrix $M$.
From a numerical point of view this is advantageous only if
truncation of $M$ is admissible. 
We defer the analysis of this question until Sec.~\ref{sec5}.
Now we concentrate on the transfer matrices themselves which,
as we shall show, possess very interesting properties.

The transfer matrix (\ref{transfer2}) exhibits the structure
of a tensor product which is conveniently displayed by switching 
to a qubit representation.
This consists in identifying the ``spatial" degree of freedom 
$q$ with the tensor product of $r$ two-level systems:
\be
|k \rangle 
\equiv
|\epsilon_0     \rangle \otimes 
|\epsilon_1     \rangle \otimes 
\ldots                  \otimes
|\epsilon_{r-1} \rangle             \, .
\ee
Here $k$ indexes the states of the $q$-basis and
$\epsilon_i=0,1$ is a label for the qubit basis states;
they are related through the binary expansion  
\be
k=
        \epsilon_0    + 
2       \epsilon_1    +    
        \ldots                 
2^{r-1} \epsilon_{r-1}  \, .
\label{binary}
\ee

\begin{figure}[htp]
\hspace{0.0cm}
\includegraphics[angle=0.0, width=8cm]{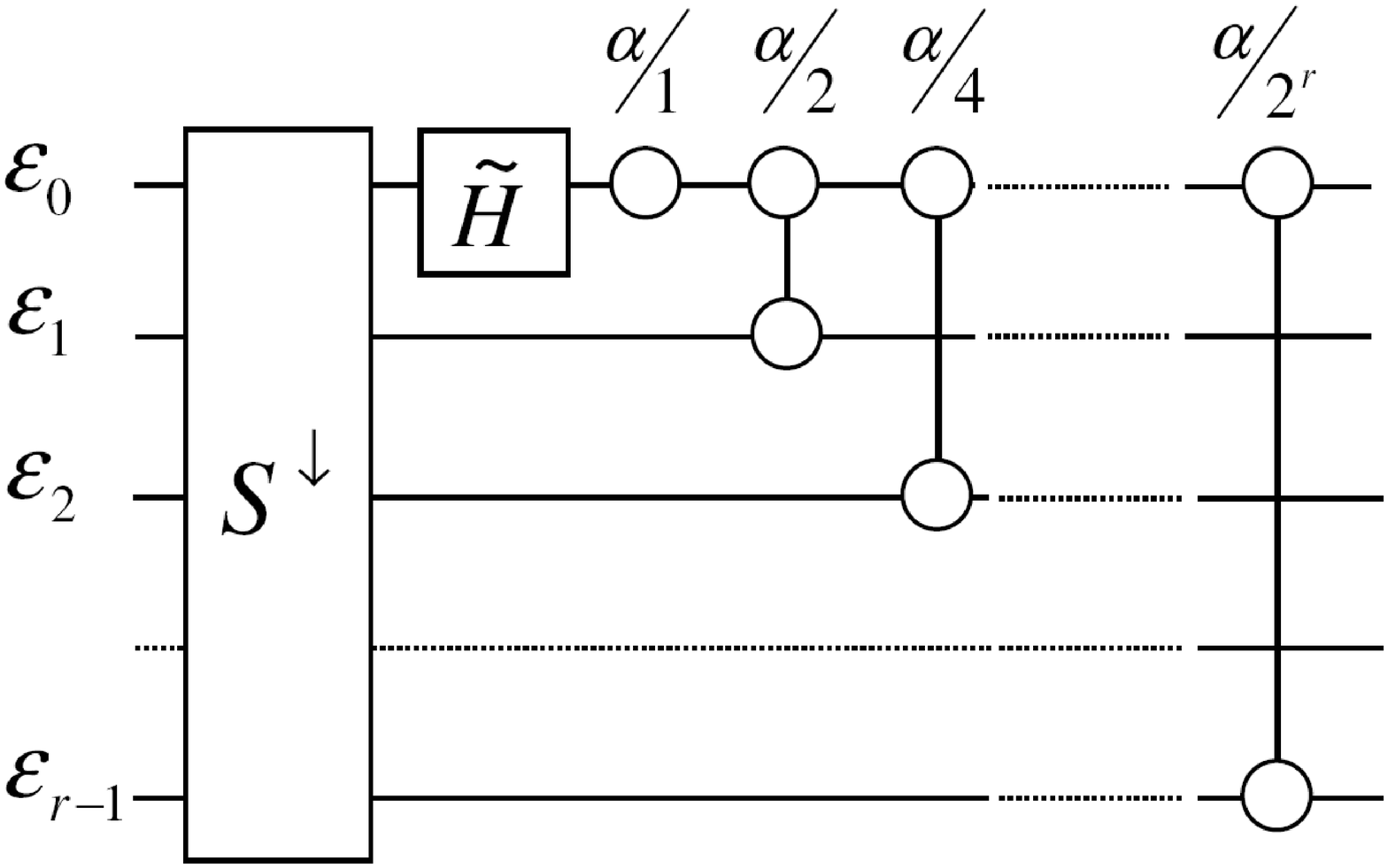}
\caption{Quantum circuit associated with the semiclassical 
trace of the $L$-th power of the baker map. 
The parameter $r \le L-1$ controls the level of truncation
($r=L-1$ corresponds to no truncation at all).
From left to right, the circuit is composed of 
a downwards shift of all qubits $(S^{\downarrow})$, 
a nonunitary one-qubit gate $\tilde{H}$,
and a sequence of symmetric phase gates.}
\label{fig2}
\end{figure}

In the qubit picture the matrix $M$ can be broken up into 
the elementary gates shown in the quantum circuit of 
Fig.~\ref{fig2}. 
%
%
%
%
The circuit acts on $r$ qubits, starting with a 
downwards cyclic shift of all qubits, i.e., 
\be
S^\downarrow 
|\epsilon_0     \rangle \otimes 
|\epsilon_1     \rangle \otimes 
\ldots                  \otimes
|\epsilon_{r-1} \rangle = 
|\epsilon_{r-1} \rangle \otimes
|\epsilon_0     \rangle \otimes 
|\epsilon_1     \rangle \otimes 
\ldots                  \otimes
|\epsilon_{r-2} \rangle \; .
\ee
A nonunitary gate acting on the first qubit follows,
which bears some resemblance with a Hadamard gate,
\be
\tilde{H}=\frac{1}{\sqrt{2}} 
\left( \begin{array}{cc}
          1 & 1 \\
          1 & x
       \end{array} \right)  \, ,
\label{tildeH}
\ee
with 
\be
x=\exp \left( \frac{2 \pi i N}{2^L-1} \right) \,.
\ee
It must be noted that, as the dimension $N$ is an 
even number, we always have $x \ne -1$, and $\tilde H$ 
is never unitary.
After the $\tilde H$ gate one finds a single qubit 
phase gate,
\be
P_{00}(\alpha)= 
\left( \begin{array}{cc}
          1 & 0 \\
          0 & {\mbox e}^{i \alpha}
       \end{array} \right) \,.
\ee
Finally one has a sequence of two-qubit phase 
gates, acting symmetrically between the first and
the $k$-th qubit, 
\be
P_{0k}(\beta)= 
\left( \begin{array}{cccc}
          1 & 0 & 0 & 0 \\
          0 & 1 & 0 & 0 \\
          0 & 0 & 1 & 0 \\
          0 & 0 & 0 & {\mbox e}^{i \beta}
       \end{array} \right) \,,
\ee
with exponentially decreasing phases, 
$\beta = \alpha/2,\alpha/4,\ldots,\alpha/2^r$.

Writing $M$ as a circuit has helped us in identifying its
basic constituents.
In particular we recognize the one-qubit gate $\tilde H$
as responsible for the deviation from unitarity.
If we substitute $\tilde H$ by a Hadamard gate, i.e.,
set $x=-1$ in (\ref{tildeH}), then the circuit of Fig.~\ref{fig2} 
acquires the typical structure of the baker family 
\cite{ermann06}:
an ``essential-baker" block 
(formed by a shift $S^\downarrow$ and a single-qubit 
Fourier transform) followed by a ``diffraction kernel"
(phase gates).


\section{Spectral Properties}
\label{sec3}

\begin{figure}[htp]
\hspace{0.0cm}
\includegraphics[angle=0.0, width=7cm]{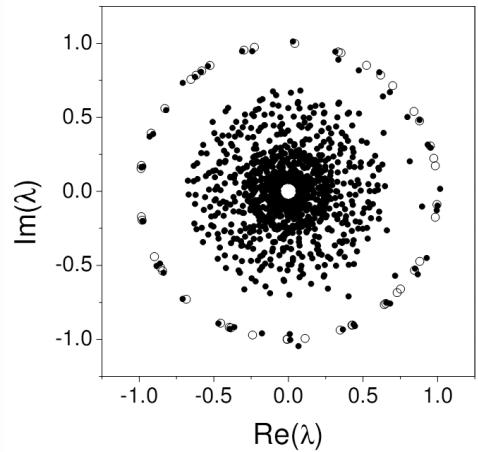}
\caption{Comparison of spectra.
Open circles correspond to the quantum baker map ($N=46$) 
and dots to the semiclassical transfer matrix for $L=11$.}
\label{fig3}
\end{figure}

Figure~\ref{fig3} displays a typical transfer-matrix 
spectrum in the case that the dimension of the transfer 
matrix is much larger than the quantum dimension, i.e., 
$2^{L-1} \gg N$.
%
%
%
%
The spectrum can be coarsely divided into three parts.
Approximately $N$ eigenvalues are localized close to
the unit circle.  
Most of the remaining ones are concentrated in 
a disk of smaller radius, and there are some transitional
eigenvalues spiraling out from the inner disk to the unit 
circle. 
The eigenvalues lying close to the unit circle can be put 
in almost one-to-one correspondence with the exact eigenvalues
of the quantum baker (\ref{bakerqq}).

\begin{figure}[htp]
\hspace{0.0cm}
\includegraphics[angle=0.0, width=8cm]{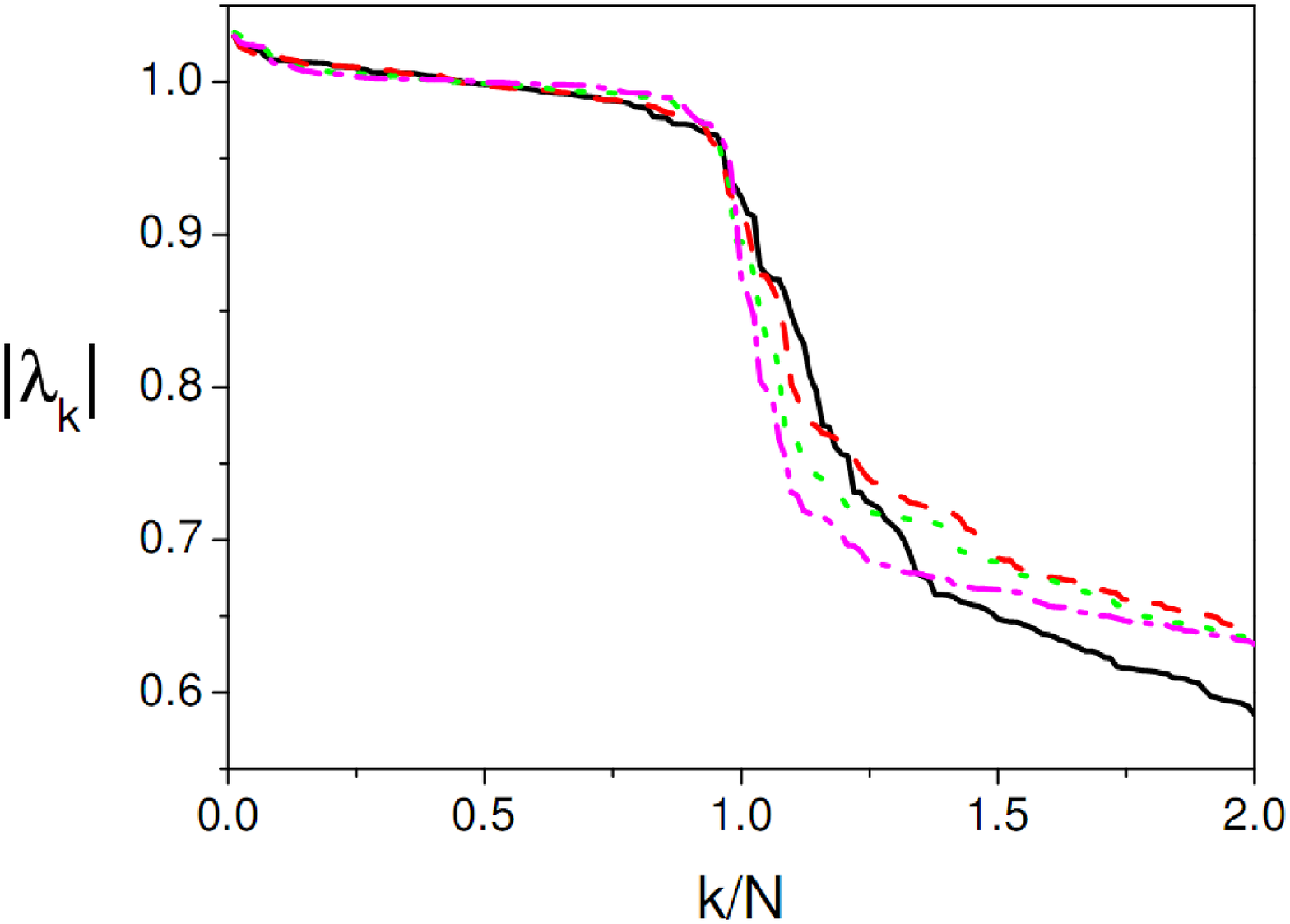}
\caption{(color online) 
Semiclassical eigenvalues (moduli) versus scaled eigenvalue 
index $k/N$ (decreasing order). 
Note the clear transition around $k/N=1$.
Parameters are $N=82$ and $L=9,10,11,12$ 
(black/solid, red/dash, green/dot, magenta/dash-dot, respectively).}
\label{fig4}
\end{figure}

Increasing $L$ while keeping $N$ fixed does not significantly 
change the almost unitary part of spectrum, but populates the 
region of small moduli (see Fig.~\ref{fig4}).
%
%
%
%
This fact, when combined with the existence of 
eigenvalues with moduli larger than one, 
leads to the conclusion that Gutzwiller's traces for the
baker are divergent in the limit $L \to \infty$ (fixed $N$). 
This result has been confirmed previously using the 
transfer-operator method \cite{argaman93,dittes94,tanner99}.

Indeed, the overall features of the transfer-matrix 
spectrum described above can also be found in the 
transfer-operator spectra discussed in 
Refs.~\cite{dittes94,tanner99}.
Looking for an explanation for this similarity, we
recall that the transfer operator is infinite-dimensional
and independent of $L$. 
Thus, one may be tempted to compare Eq.~(\ref{Wqq}) with 
the infinite-$L$ limit of the transfer matrix 
(\ref{transfer2}).
In this limit, the transfer matrix becomes an integral
kernel, indices becoming continuous variables:
\be
\frac{k}{2^r-1} \to q \, , \qquad 0 \le q \le 1 \, .
\ee
A careful calculation verifies that the transfer
matrix tends to the transfer operator, i.e.,
\be
\lim_{L \to \infty} M_{kk^\prime} = W(q,q^\prime) \, ,
\ee
with $W(q,q^\prime)$ precisely that given in 
Eq.~(\ref{Wqq})!
So, the transfer operator is formally recovered as 
the continuum limit of the transfer matrix.

\section{Truncation Schemes} 
\label{sec4}

Here we discuss the numerical calculation of long-time 
Gutzwiller traces using the transfer-matrix formula 
(\ref{transfer}). 
The most natural procedure for calculating ${\rm tr} M^L$ 
requires the diagonalization of $M$ to obtain its 
eigenvalues. 
This scheme is limited to very small values of $L$,
e.g., to $L \lesssim 15$ in a personal computer.
So, truncation becomes essential.

The error introduced in truncating the interactions
to $r$ neighbors can be roughly estimated from 
Eq.~(\ref{truncation}) and the basic definitions
(\ref{gutzwiller},\ref{action},\ref{matrixA}).
Such a truncation produces an error in the actions 
(\ref{action}) approximately given by
\be
\delta {\cal S} \sim
\frac{1}{2} \, 
\epsilon^t \,
 \frac{S^r}{2^r} \, \epsilon \; . 
\ee 
For typical vectors $\epsilon$, containing 
randomly distributed 0/1 bits, we have 
\be
\epsilon^t \, S^r \, \epsilon \sim \frac{L}{4} \, . 
\ee 
This implies that the corresponding error in the 
Gutzwiller phases amounts to
\be
\delta \phi \equiv \delta \left( 2 \pi N {\cal S}  \right) 
\sim 
\frac{N L}{2^r} \; . 
\ee 
Finally, if we assume that this is the error committed 
in most phases in the Gutzwiller sum (\ref{gutzwiller}), 
then $\delta \phi$ amounts to the relative error in the 
semiclassical traces, for
\be
\delta \sum e^{i \phi_k} = 
\sum e^{i \phi_k} \delta \phi_k \approx
\delta \phi \sum e^{i \phi_k}  \; . 
\ee 
Thus, we arrive at the following condition for the 
validity of the truncation to $r$ neighbors:  
\be
\frac{2^r}{N L} \equiv f \gg 1 \, .
\label{bound}
\ee

However, even after truncating the transfer matrix, 
we may need to use values of $r$ which make
diagonalization prohibitive, e.g., for $N=50$, $L=1000$.
Remarkably there is an alternative to diagonalization
which arises from the sparsity of $M$.
For, even if the special type of sparsity $M$ exhibits is 
not helpful in speeding-up its diagonalization, it permits 
us to implement an alternative, much more efficient 
method for calculating the traces of $M$.

Consider the following identity:
\be
{\rm tr} M_r^L =  2^r \, 
                 \overline{
                 \langle \psi \left| M_r^L
                 \right| \psi \rangle    
                          }   \, ,
\label{CUE}
\ee
where overline means average over random complex 
vectors $|\psi \rangle$ uniformly distributed over 
the corresponding unit sphere \cite{mehta}. 
(The notation $M_r$ has been used to indicate the 
truncation of $M$ to $r$ qubits.)
The idea behind (\ref{CUE}) is to calculate 
$\langle \psi | M_r^L | \psi \rangle$
by applying iteratively the matrix $M_r$ to 
$|\psi \rangle$, and then averaging over 
$|\psi \rangle$.
This method allows one, in principle, to consider
$r$ as large as 20. 

Table~\ref{tab1} displays some examples.
\begin{table}[ht]
\vspace{1pc}
\begin{center}
\begin{tabular}{|c|c|c|l|l|c|}   
\hline
~~~~         &
$\;L\;$      &  
$\;\;r\;\;$  &  
~~TMM        & 
~~TOM        & 
~exact~        
\\ \hline \hline
a & 20  &  10  &  $\;3.73[5]    $ &  $\;3.87[6]    $  & 3.85   \\ \hline
b & 20  &  11  &  $\;3.90[7]    $ &  $\;3.87[6]    $  & 3.85   \\ \hline
c & 20  &  12  &  $\;3.85[3]    $ &  $\;3.87[6]    $  & 3.85   \\ \hline
d & 20  &  13  &  $\;3.84[5]    $ &  $\;3.87[6]    $  & 3.85   \\ \hline
e & 20  &  14  &  $\;3.89[5]    $ &  $\;3.87[6]    $  & 3.85   \\ \hline
f & 40  &  13  &  $\;2.04[2]10^1$ &  $\;1.99[1]10^1$  &   ?    \\ \hline
g & 60  &  13  &  $\;1.00[5]10^1$ &  $\;1.05[5]10^1$  &   ?    \\ \hline
h & 80  &  13  &  $\;1.30[2]10^2$ &  $\;1.24[1]10^2$  &   ?    \\ \hline
i & 160 &  13  &  $\;1.00[1]10^4$ &  $\;9.0[3] 10^3$  &   ?    \\ \hline
\end{tabular}
\end{center}
\caption{%
Gutzwiller's traces (absolute value).
We compare the transfer-matrix method (TMM) with 
the transfer-operator method (TOM) and exact results.
In all cases the quantum dimension is $N=24$.
The square brackets contain the estimated error in the 
least significative figure [e.g., the notation
$1.30[2]10^2$ stands for $(1.30\pm0.02)\times10^2$].
The values in the TMM column are the result of 
averaging over more than $10^5$ random states; 
the errors are statistical.
In the TOM case the errors correspond to the variation
of the results when the size of the Fourier basis is
varied from 149 to 201 \cite{dittes94,tanner99}.}
\label{tab1}
\end{table}
The first rows (a-e) illustrate the improvement of the 
results as the truncation level $r$ is increased, for
fixed $(N,L)$.
Agreement with the exact results is met, within
the specified statistical errors, for $r \ge 12$, 
corresponding to $f \gtrsim 4$.
In rows (f-i) we consider $(N,r)$ fixed and large values 
of $L$ which can in no way be reached by direct computation 
of the periodic orbit summation, so that exact results are 
not known for such $L$.
In these cases we compare with the approximate results 
obtained using the transfer-operator method 
\cite{dittes94,tanner99}.

We see good agreement for $L \lesssim 80$, i.e.,
we find the condition $f \gtrsim 4$ again.

So, we have checked that the transfer-matrix method 
works satisfactorily in the parameter regime specified
by (\ref{bound}).

\section{Divergence of long-time traces} 
\label{sec5}

For chaotic systems, the Gutzwiller trace formula 
is only the first term of an expansion in powers of 
$\hbar$.
Because of this, the semiclassical eigenvalues 
obtained from the Gutwiller traces deviate in general 
from the unit circle in the complex plane.
For typical chaotic systems it is expected 
that the distance of the largest semiclassical eigenvalue 
to the 
unit circle scales like $\hbar$, i.e., $1/N$ for 
quantum maps \cite{keating94}.
In the case of the baker map, an anomalous behavior
$N^{-1/2}$ is observed which is due to diffraction 
effects originating in the discontinuities of the 
mapping \cite{dittes94,tanner99}. 
In any case, this lack of unitarity makes Gutzwiller's 
traces exponentially growing with $L$, a behavior that 
may be guessed from Table~\ref{tab1}.

The following question arises naturally: 
can the $N^{-1/2}$ law be associated with the 
particular {\em structure} of the baker transfer matrix?
In the circuit representation (Fig.~\ref{fig2})
nonunitarity arises from the single-qubit gate 
$\tilde H$.
So, one may ask: what is typically the largest eigenvalue 
(in modulus) of a map obtained by tensoring $\tilde H$ 
with a generic $N$-dimensional unitary gate $U$?

On the other side, the $q$-representation (\ref{transfer2}) 
offers a different point of view of the nonunitarity: 
the transfer matrix can be split as 
$M=(U_1+U_2)/\sqrt{2}$, with $U_i$
unitary matrices of dimension $N$ [see Eq.~(\ref{t3})].
How does the leading eigenvalue of such a matrix $M$ scale with 
$N$?

In order to answer the questions above we resorted
to numerical calculations. 
First we analyzed the matrices $H' \otimes U$ with 
$U$ a random matrix belonging to the Circular Unitary
Ensemble (CUE). 
The matrix $H'$ acts on the least significant qubit and
is defined by Eq.~(\ref{tildeH}) with 
$x=1$, i.e., $H'$ is chosen to be the large-$L$ limit 
of $\tilde H$.
Then we considered the ensemble $(U_1+U_2)/\sqrt{2}$
with $U_i$ independent random matrices belonging to
CUE  \cite{gorlich04}. 
In order to check any possible ensemble dependence
we also calculated the case $H' \otimes U$ with $U$ in 
the Circular Orthogonal Ensemble (COE).

\begin{figure}[htp]
\hspace{0.0cm}
\includegraphics[angle=0.0, width=8cm]{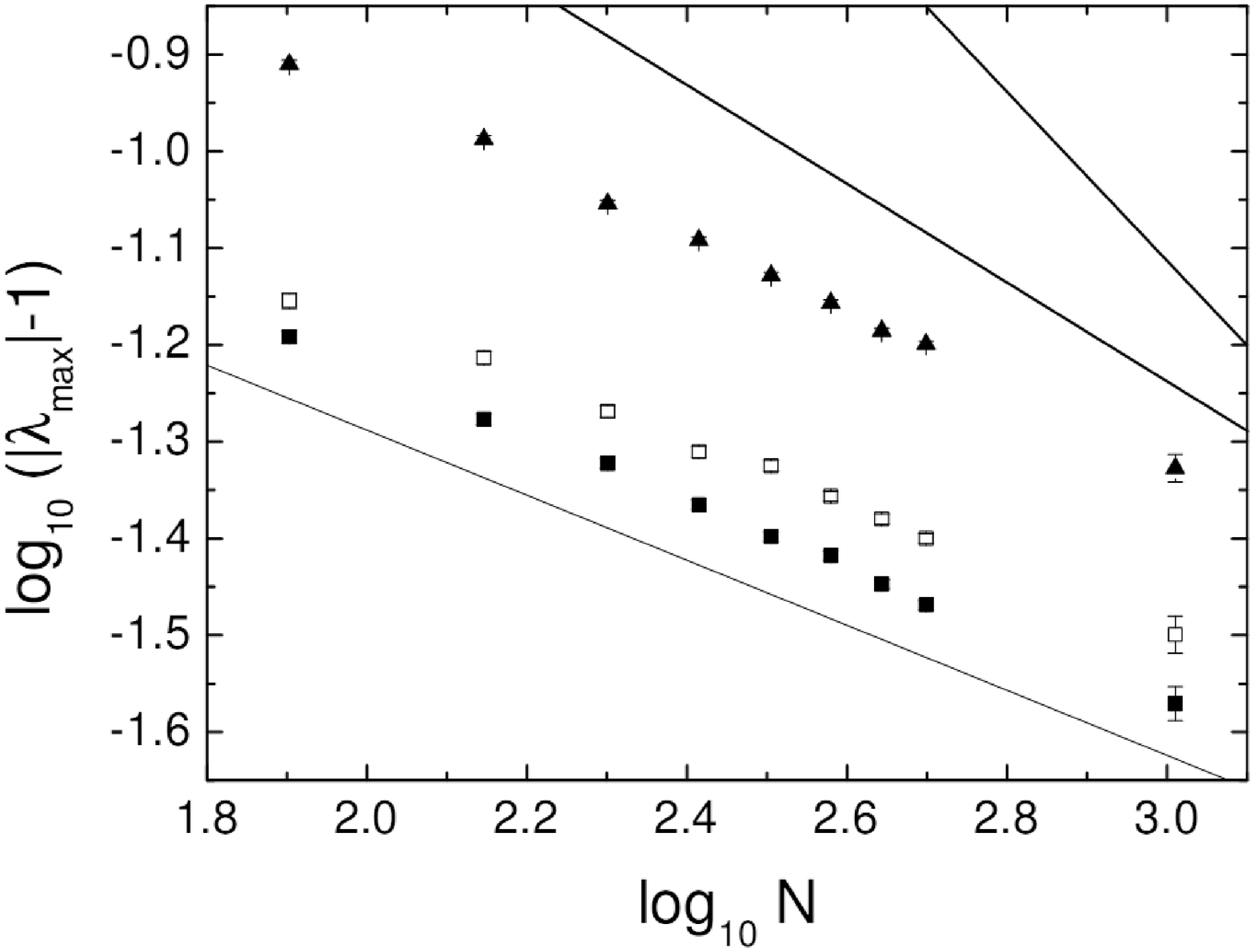}
\caption{%
Average distance of the largest eigenvalue to the unit 
circle versus dimension for three ensembles of nonunitary 
random matrices.
Triangles, squares, and circles correspond to 
$H' \otimes \rm{COE}$, 
$(\rm{CUE}+\rm{CUE})/\sqrt{2}$ 
(normalized sum of two independent CUE matrices), 
and 
$H' \otimes \rm{CUE}$, respectively. 
See text for the definition of the single-qubit 
nonunitary gate $H'$.
Error bars correspond to statistical errors.
Straight lines representing decay laws  
$N^{-1/3}$, $N^{-1/2}$, and $N^{-1}$ have been drawn 
for reference.}
\label{fig5}
\end{figure}

Our results are exhibited in Fig.~\ref{fig5}.
%
%
In the three considered cases we observe very similar 
decay laws.
These decays are much closer to the diffractive scaling 
$N^{-1/2}$ than to the universal semiclassical behavior 
$N^{-1}$, meaning that the random-matrix modeling
has partially captured the essence of the semiclassical
baker. 
However, there is still some noticeable departure from 
the $N^{-1/2}$ decay (rather, our numerical results 
seem to follow a $N^{-1/3}$ law).
Thus, we conclude that the minimum-information
models we have constructed still have to be complemented
with some ingredient to properly describe the  
$N^{-1/2}$ baker scaling.
Further research is necessary to discover what such an 
additional information should be (e.g., some particular 
correlation between $U_1$ and $U_2$?).

\section{Reflected baker map} 
\label{sec6}

The quantum baker map $B$ is invariant under the parity 
symmetry $R$, its action on the $q$-basis being just 
$R\,|k \rangle = |N-k \rangle$.
This is the quantum counterpart of the classical reflection
symmetry ${\cal R}:(q,p) \rightarrow (1-q,1-p)$. 
When analyzing spectral properties it is convenient to
separate eigenstates and eigenvalues of $B$ according to 
their parity.
Thus, one is led to consider the parity-projected bakers
$B_{\pm}=(B \pm B R)/2$. 
In the semiclassical domain the traces of $B_\pm$ can be 
approximated by \cite{saraceno94}
\be
{\rm tr} \; B^L_{\pm} 
\approx \frac{1}{2} \,\, \left(
\frac{2^{L/2}}{2^{L}-1} 
\sum_{\epsilon}  
e^{2\pi i N S(\epsilon)} ~ 
\pm ~ \\
\frac{2^{L/2}}{2^{L}+1} 
\sum_{\epsilon}  
e^{2\pi i N S'(\epsilon)} 
\right) \, .
\label{gutzsim}
\ee
The first summation above correspond to the baker traces 
(\ref{gutzwiller}). 
In the second sum $S'(\epsilon)$ stands for half the action 
[Eq.~(\ref{action})] of a periodic orbit of length $2L$ with 
symbolic code 
$\epsilon=
(\epsilon_0,\epsilon_1,\ldots,\epsilon_{L-1},
\bar{\epsilon_0},\bar{\epsilon_1},\ldots,
\bar{\epsilon_{L-1}})^t$, 
where
$\bar{\epsilon_i}=1-\epsilon_i$. 
These are precisely the orbits invariant 
under the parity transformation. 

We shall exhibit the transfer
matrix associated to the reflected traces, pointing out
the most important features. 
Following the same procedure as in Sec.~\ref{sec2}
we constructed a transfer matrix $M'$ such that
\be
\sum_{\epsilon}  e^{2\pi i N S'(\epsilon)} 
= \tr M'^{\, 2L} \, . 
\ee 
After eliminating the null subspace that appears as a 
consequence of considering just the parity-invariant 
trajectories of length $2L$, the matrix
$M'$ is reduced to dimension $2^L \times 2^L$:
\be
M' =
\left( \begin{array}{cccccccc}
          0 & 1 & 0 & 0 & \cdots & \cdots & 0 & 0  \\
          0 & 0 & 0 & 1 &    0   & \cdots & 0 & 0  \\
   \vdots & \vdots & \vdots & \vdots & 
   \vdots & \ddots & \vdots & \vdots               \\
          0 & 0 & 0 & 0 & \cdots & \cdots & 0 & 1  \\
      a_0 & 0 & 0 & 0 & \cdots & \cdots & 0 & 0  \\
      0 & 0 & a_1 & 0 & \cdots & \cdots & 0 & 0  \\
   \vdots & \vdots & 0 & 0 & \cdots & 
   \ddots & \vdots & \vdots  \\
          0 & \cdots & 0 & 0 & \cdots & 
              \cdots & a_{2^{L-1}-1} & 0
       \end{array} \right) \,,
\label{t3}
\ee
where the matrix elements $a_k$ are given by
\be
a_k=\exp \left[ i \alpha'
\left( 1+ \frac{\tilde{k}}{2^{2L-1}} \right) \right] \,,
\ee
with 
\be
\alpha'= \frac{\pi N \, 2^{2L-1} }{2^{2L}-1}  \,.
\ee
The last undefined ingredient is the integer $\tilde k$.
It is best expressed in terms of the binary digits of the
integer $k$ [see Eq.~(\ref{binary})]. 
If 
$k=\epsilon_{2^{L-1}-1} \ldots \epsilon_1 \epsilon_0 $, 
then 
\be
\tilde k=
\epsilon_{2^{L-1}-1} 
\ldots 
\epsilon_1 
\epsilon_0 
\, 0 \,
\overline{\epsilon_{2^{L-1}-1}} 
\ldots 
\bar{\epsilon_1} 
\bar{\epsilon_0} \, .
\ee

\begin{figure}[htp]
\hspace{0.0cm}
\includegraphics[angle=0.0, width=8cm]{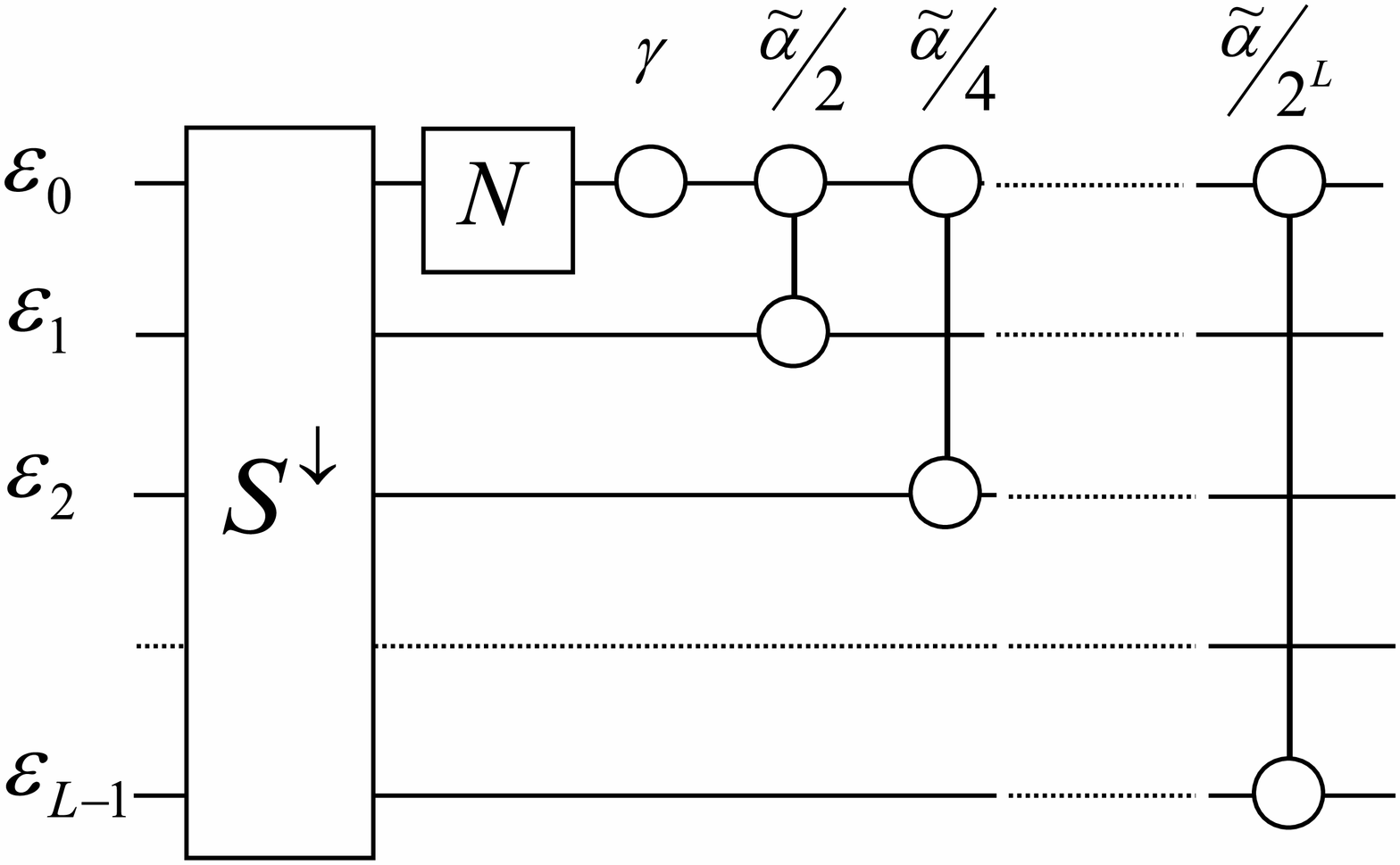}
\caption{Quantum circuit associated with the semiclassical 
traces of the parity-reflected baker map. 
From left to right, the circuit is composed of 
a downwards shift of all qubits $(S^{\downarrow})$, 
a NOT gate ($N$), a one-qubit gate, and a sequence of $L-1$ 
two qubit phase gates. All gates are unitary.}
\label{fig6}
\end{figure}

The main difference between the baker's transfer matrix 
(\ref{transfer2}) and the matrix above, is that $M'$ is
exactly unitary (thus, its spectrum lies on the unit
circle). 
Except for this fact, both $M$ and $M'$ are structurally
very similar. 
Perhaps this can be better appreciated by looking at the
quantum circuit for $M'$ in Fig.~\ref{fig6}.
%
%
The circuit acts on $L$ qubits (without truncation). 
It starts like the baker's, with a downwards shift of all qubits. 
Then we have a NOT gate acting on the first qubit given by
\be
N= 
\left( \begin{array}{cc}
          0 & 1 \\
          1 & 0
       \end{array} \right) \, .
\ee
A phase gate $P_{00}(\gamma)$ follows that acts on the first qubit, 
with 
\be
\gamma = \alpha' \left( 1 + \frac{2^{L-1}-1}{2^{2L-1}} 
                  \right) \,.
\ee
Finally there is a sequence of $L-1$ symmetrical two-qubit phase 
gates $P_{0k}(\beta)$ acting between the first and the $k$-th qubit
with exponentially decreasing phases, 
$\beta = 
\tilde{\alpha}/2,
\tilde{\alpha}/4,
\ldots,
\tilde{\alpha}/2^L$, 
where 
$\tilde{\alpha}=\alpha' (2^L-1)/2^L$.

A complete study of the spectral properties of the 
reflected-baker transfer matrix, 
truncation schemes, etc, is deferred to a future 
publication because, as $M'$ is unitary, such analyses 
would take us in directions very different than those 
followed in the case of the baker map.
However, we would like to anticipate one result 
which gives a hint about the nature of the spectrum of 
$M'$.

Consider the $2L$-th power of the transfer matrix $M'$, 
i.e., the $2L$-th iteration of the circuit above.
We verified that the shifts and NOT gates cancel out
and only phase gates remain.
This means that $M'^{\, 2L}$ is a diagonal matrix 
(given that the phase gates are diagonal).
It turns out that the diagonal elements 
are exactly the Gutwiller phases $\exp (iS'/\hbar)$ 
in Eq.~(\ref{gutzsim})
(we tested this numerically for some cases).
Thus, the spectrum of $M'$ is formed by roots of the 
Gutzwiller phases!

\section{Conclusions} 
\label{sec7}

We presented a study of the transfer matrix approach to
the semiclassical traces of the baker map and its
reflected version.
We found that the transfer matrices admit a tensor product
decomposition leading to simple circuit representations, 
similar to those obtained by Schack and Caves for the 
quantum baker.
Remarkably, in the case of the baker, the corresponding 
circuit clearly isolates the source of nonunitarity of the 
semiclassical traces in the form of a one-qubit Hadamard-like 
gate.
Given that both exact and semiclassical bakers can now be 
written as circuits, this representation appears as a 
promising tool for studying the corrections to the Gutzwiller 
trace formula.

In the case of the baker, spectral properties were analyzed, 
showing that there exists a close similitude with the transfer 
operator of Ref.~\cite{dittes94}.
In fact, we proved that the transfer operator arises as 
a suitable asymptotic limit of the transfer matrix 
formalism.
Truncation schemes were discussed which
permit the numerical calculation of long-time traces well 
inside a domain where direct summation of the Gutzwiller 
formula is impossible.

We would like to conclude by mentioning a very exciting
(though rather speculative) possibility.
It is tempting to think of the transfer matrices 
as maps for which the Gutzwiller formula is exact.
Admittedly this point of view has some limitations
because there is a map for each value of $L$. 
Even so, our results suggest that we may be not far 
from finding a {\em close relative} to the baker map 
having an exact periodic-orbit trace formula. 
The search for such a map constitutes a very attractive
and challenging project which, however, exceeds the 
scope of the present paper.


\section*{Acknowledgments}

We thank 
M. Saraceno, 
A. M. Ozorio de Almeida,
and
E. Vergini  
for many useful comments.
Partial financial support from agencies 
CNPq, FAPERJ (Brazil), and CONICET (Argentina) 
is acknowledged.
G. C. is a researcher of CONICET. 

\vspace{3pc}


\end{document}